\documentclass{mn2e}
\usepackage{graphicx}
\usepackage{amsbsy}
\usepackage{amsmath}
\usepackage{breqn}
\usepackage{epsf}
\usepackage{paralist}
\usepackage{upgreek}
\usepackage{tipa}
\usepackage[inline]{enumitem}
\usepackage{arydshln}
\usepackage{amssymb}
\usepackage{pifont}
\usepackage{epstopdf}

\newcommand{\beq}{\begin{equation}}
\newcommand{\eeq}{\end{equation}}

\DeclareMathAlphabet{\mathsfsl}{OT1}{cmss}{bx}{sl}
\SetMathAlphabet{\mathsfsl}{bold}{OT1}{cmss}{bx}{sl}

\newcommand{\RN}[1]{%
  \textup{\uppercase\expandafter{\romannumeral#1}}%
}

\def\mean#1{\left< #1 \right>}
\usepackage{color}        

\begin{document}

\title[Observing drift velocities in cores]
{Can we observe the ion-neutral drift velocity in prestellar cores?}

\author[Tritsis et al.]
  {Aris~Tritsis$^{1, 2}$\thanks{E-mail: aris.tritsis@epfl.ch}, Shantanu Basu$^{1}$, and Christoph Federrath$^{3}$ \\
    $^1$Department of Physics and Astronomy, University of Western Ontario, London, ON N6A 3K7, Canada \\
    $^2$Institute of Physics, Laboratory of Astrophysics, Ecole Polytechnique F\'ed\'erale de Lausanne (EPFL), \\ Observatoire de Sauverny, 1290, Versoix, Switzerland \\
    $^3$Research School of Astronomy and Astrophysics, Australian National University, Canberra, ACT 2611, Australia} 
\maketitle 

\begin{abstract}

Given the low ionization fraction of molecular clouds, ambipolar diffusion is thought to be an integral process in star formation. However, chemical and radiative-transfer effects, observational challenges, and the fact that the ion-neutral drift velocity is inherently very small render a definite detection of ambipolar diffusion extremely non-trivial. Here, we study the ion-neutral drift velocity in a suite of chemodynamical, non-ideal magnetohydrodynamic (MHD), two-dimensional axisymmetric simulations of prestellar cores where we alter the temperature, cosmic-ray ionization rate, visual extinction, mass-to-flux ratio, and chemical evolution. Subsequently, we perform a number of non-local thermodynamic equilibrium (non-LTE) radiative-transfer calculations considering various idealized and non-idealized scenarios in order to assess which factor (chemistry, radiative transfer and/or observational difficulties) is the most challenging to overcome in our efforts to detect the ion-neutral drift velocity. We find that temperature has a significant effect in the amplitude of the drift velocity with the coldest modelled cores (\textit{T} = 6 K) exhibiting drift velocities comparable to the sound speed. Against expectations, we find that in idealized scenarios (where two species are perfectly chemically co-evolving) the drift velocity ``survives" radiative-transfer effects and can in principle be observed. However, we find that observational challenges and chemical effects can significantly hinder our view of the ion-neutral drift velocity. Finally, we propose that $\rm{HCN}$ and $\rm{HCNH^+}$, being chemically co-evolving, could be used in future observational studies aiming to measure the ion-neutral drift velocity.

\end{abstract}


\section{Introduction}\label{intro}

Non-ideal magnetohydrodynamic (MHD) effects and especially ambipolar diffusion have been shown to have an integral role in the formation of stars (e.g. Mouschovias \& Ciolek 1999). Firstly, they allow the redistribution of magnetic flux within molecular clouds, thereby allowing their self gravity to overcome the support from interstellar magnetic fields so that they can gravitationally collapse (Basu et al. 2009; Kunz \& Mouschovias 2010; Tassis et al. 2012a; Tritsis et al. 2022). Secondly, the removal of magnetic flux during the prestellar and protostellar phase is necessary to explain the so called ``magnetic-flux problem" (e.g. Tassis \& Mouschovias 2005; Tsukamoto et al. 2015). Thirdly, non-ideal MHD effects reduce the effectiveness of magnetic braking, thus allowing the formation of rotationally-supported discs (see e.g. Wurster \& Lewis 2020 and references therein).

Despite the importance of ambipolar diffusion, a definite detection of its observational signatures is yet to be confirmed. This is not due lack of trying but rather because observing the ion-neutral drift velocity is a very challenging task. Firstly, the ion-neutral drift velocity theoretically predicted from numerical simulations is typically of the order of $\sim$10-50\% of the sound speed ($c_{\rm{s}}\sim0.2~\rm{km~s^{-1}}$ e.g. Desch \& Mouschovias 2001; Tritsis et al. 2022). In comparison, the spectral resolution of most modern radio-telescopes is of the order of 0.1--0.2 $\rm{km~s^{-1}}$. To make matters worse, the observed charged and neutral molecular species need to be chemically co-evolving (Tassis et al. 2012b). In the opposite scenario, any differences in the linewidth between the observed spectra of the charged and neutral species can be simply attributed to the fact that the two molecules probe different regions of the core which are collapsing at different rates. Thirdly, even in the limit where the spectral resolution of a radio-telescope can accurately probe subsonic motions and the ion and neutral species are perfectly co-evolving, radiative-transfer effects can complicate the picture so that any information regarding their velocity difference is lost. 

Indirect evidence/indications for ambipolar diffusion in molecular clouds have been found more than three decades ago (Myers \& Goodman 1988). More recently, Li \& Houde (2008) used $\rm{HCN}$ (\textit{J} = 4$\rightarrow$3) and $\rm{HCO^+}$ (\textit{J} = 4$\rightarrow$3) observations from the M17 star-forming region and determined that the neutral species exhibits higher velocity dispersion than the ion. Similarly, Hezareh et al. (2010) observed the $\rm{H^{13}CN}$ and $\rm{H^{13}CO^+}$ (\textit{J} = 4$\rightarrow$3) transitions towards the DR21(OH) star-forming region and arrived at the same conclusion (see also Tang et al. 2018). On the other hand, Pineda et al. (2021) performed $\rm{NH_3}$ and $\rm{N_2H^+}$ observations towards the dense core Barnard 5 and found that the velocity dispersion of the charged species was higher than that of the neutral. They interpreted their results as evidence of penetration of hydromagnetic waves into the densest regions of the core. At smaller scales, Yen et al. (2018) observed the $\rm{H^{13}CO^+}$ (\textit{J} = 3$\rightarrow$2) and $\rm{C^{18}O}$ (\textit{J} = 2$\rightarrow$1) transitions towards the Class 0 protostar B335 with ALMA and placed an upper limit on the ion-neutral drift velocity of 0.35 $\rm{km~s^{-1}}$ at a radius of 100 au. On the theoretical front, Lankhaar \& Vlemmings (2020) recently proposed that ambipolar diffusion would give rise to a preferred direction of collisions between charged species and $\rm{H_2}$ which would then lead to the emission of charged species to be linearly polarized. For typical values of the ion-neutral drift velocity, under local thermodynamic equilibrium, they predicted a polarization fraction of the order of $\sim$1--3\%.

Despite the importance of such observations and new techniques, a systematic theoretical study aiming at exploring to what degree such spectra can be used to probe ambipolar diffusion is still lacking\footnote{In a novel study, Yin et al. (2021) recently produced a number of mock observations of line spectra for various molecules from super- and sub-critical models of molecular clouds. However, they did not address whether gravitationally-initiated ambipolar diffusion can be probed on the basis of such spectra.}. Here, we aim to bridge this gap by performing a suite of non-ideal MHD chemodynamical simulations of prestellar cores. We then design a number of radiative-transfer numerical experiments to explore to what degree spectral-line observations can be used to probe ambipolar diffusion.

Non-ideal MHD chemodynamical simulations pose a significant numerical challenge with most studies either following a limited number of molecular species (e.g. Kunz \& Mouschovias 2009) or others addopting the ``thin-disk" approximation (Tassis et al. 2012a), in order to save computational time. Tritsis et al. (2022) performed two-dimensional (2D) axisymmetric simulations with two very extended gas-grain chemical networks, consisting of $\sim$300 and 115 species, and studied the effect of deuterium chemistry, grain distribution (or absence of grains) and that of the cosmic-ray ionization rate on non-ideal MHD effects. They found that deuterium chemistry has a significant effect on non-ideal MHD effects with $\rm{D_3^+}$ and $\rm{HD_2^+}$ being the main carriers of the conductivities at high densities ($n_{\rm{H_2}}\gtrsim 10^6~\rm{cm^{-3}}$).

This paper is organized as follows: In section \S~\ref{dyna} we describe our chemodynamical simulations and present our theoretical results. In \S~\ref{RT} we present the details of our radiative-transfer calculations and describe our strategy to methodically assess which is the dominant factor affecting our view of the ion-neutral drift velocity. In \S~\ref{ChemEvolSec} we perform a search for chemically co-evolving molecules and finally, in \S~\ref{discuss} we summarize our results and conclude.

\section{Chemodynamical Simulations}\label{dyna}

\subsection{Numerical setup}\label{dynanumer}

We follow Tritsis et al. (2022) to perform a total of eleven 2D axisymmetric, chemodynamical, non-ideal magnetohydrodynamic (MHD) simulations of collapsing prestellar cores using a modified version of the \textsc{FLASH} astrophysical code (Fryxell et al. 2000; Dubey et al. 2008; Tritsis et al. 2022). We explore a large part of the parameter space. Specifically, we alter the temperature, cosmic-ray ionization rate, visual extinction and mass-to-flux ratio. We also perform an additional simulation where we leave the chemistry to evolve for 2 Myr before switching on gravity. The physical parameters of each of our simulations are listed in Table~\ref{simstable}.

We refer the reader to Tritsis et al. (2022) for a detailed description of the basic equations and our numerical setup. Below we summarize the main properties of our simulated cores. The initial density in all of our simulations was set to $n_{\rm{H_2}} = 300 ~\rm{cm^{-3}}$ and we alter the strength of the magnetic field to explore different values of the mass-to-flux ratio. Additionally, all of our simulations are isothermal and all the velocity components are initially set to zero. The visual extinction which affects the reaction rates of photo-related processes in our chemical model, does not depend on the density structure of the core, but is instead set equal to the value quoted in Table~\ref{simstable} everywhere in the simulation region. Therefore, we essentially assume that the cores modelled are well embedded inside a molecular cloud. A more physical approach where the visual extinction is calculated ``on the fly" based on the density structure of the core will be explored in a future study. The radius and half-height of our simulation region are both set to 0.75 pc such that the mass of the core is $\sim$47 $\rm{M_\odot}$. We use periodic boundary conditions in the $z$ direction, a diode (i.e. open) boundary condition at $r = R$ and an axisymmetric boundary condition (i.e. reflecting -- zero flux boundary) at $r = 0$. We use an adaptive mesh refinement (AMR) grid of initial size 32$\times$64 grid points and six levels of refinement such that the size of the smallest cell is $\sim$150 au. Our grid is refined based on the density and the $z$-component of the magnetic field using the modified second derivative criterion presented in L{\"o}hner (1987). With this refinement criterion, the Truelove condition (Truelove et al. 1997) is satisfied at all times.

Our chemical network consists of 115 species and $\sim$1650 chemical reactions and the only molecular species in our initial conditions is $\rm{H_2}$. The initial elemental abundances (relative to total hydrogen nuclei) used in our network are given in Table 1 of Tritsis et al. (2022). Out of the 115 species, 37 are in the dust phase. Even though we model the chemical evolution of dust species, we do not take grains into account when calculating the resistivities. Tritsis et al. (2022) found that, for a core with the same physical conditions as in $\texttt{M/$\Phi$0.5\_$\zeta/\zeta_01$\_$A_v$10\_T10}$ (see Table~\ref{simstable}), the collapse of the cloud is delayed by $\sim$1.4 Myr when grains are taken into account when computing the resistivities since negatively charged grains can also couple to the magnetic field. However, the resulting chemical composition of the core is almost identical to the case where grains are excluded from the calculation of the resistivities (see Figure 10 in Tritsis et al. 2022). As a result, given the minimal differences in the abundance of most commonly observed molecular species and the fact that the inclusion of grains would lead to a significant increase in the computational cost (especially considering that we explore a large parameter space), we chose to exclude grains from the calculation of the resistivities in the present study.

\begin{table}
\begin{center}
\begin{tabular}{c c c c c c}
\hline\hline
& Model name & M/$\Upphi$ & $\zeta$/$\zeta_0$ & $A_v$ & T [K] \\ 
 \hline
1 & \texttt{M/$\Phi$0.5\_$\zeta/\zeta_01$\_$A_v$10\_T10} & 0.5 & 1 & 10 & 10 \\  
2 & \texttt{M/$\Phi$0.75\_$\zeta/\zeta_01$\_$A_v$10\_T10} & 0.75 & 1 & 10 & 10 \\  
3 & \texttt{M/$\Phi$0.25\_$\zeta/\zeta_01$\_$A_v$10\_T10} & 0.25 & 1 & 10 & 10 \\  
4 & \texttt{M/$\Phi$2.6\_$\zeta/\zeta_01$\_$A_v$10\_T10} & 2.6 & 1 & 10 & 10 \\  
 \hline
5 & \texttt{M/$\Phi$0.5\_$\zeta/\zeta_00.5$\_$A_v$10\_T10} & 0.5 & 0.5 & 10 & 10 \\  
6 & \texttt{M/$\Phi$0.5\_$\zeta/\zeta_02$\_$A_v$10\_T10} & 0.5 & 2 & 10 & 10 \\  
 \hline
7 & \texttt{M/$\Phi$0.5\_$\zeta/\zeta_01$\_$A_v$5\_T10} & 0.5 & 1 & 5 & 10 \\  
8 & \texttt{M/$\Phi$0.5\_$\zeta/\zeta_01$\_$A_v$20\_T10} & 0.5 & 1 & 20 & 10 \\  
 \hline
9 & \texttt{M/$\Phi$0.5\_$\zeta/\zeta_01$\_$A_v$10\_T6} & 0.5 & 1 & 10 & 6 \\  
10 & \texttt{M/$\Phi$0.5\_$\zeta/\zeta_01$\_$A_v$10\_T8} & 0.5 & 1 & 10 & 8 \\  
 \hline
11 & \texttt{M/$\Phi$0.5\_$\zeta/\zeta_01$\_$A_v$10\_T10\_dC} & 0.5 & 1 & 10 & 10 \\  
\hline\hline
\end{tabular}
\end{center}
\caption{\label{simstable} Physical parameters of the chemodynamical simulations performed. We take the standard value of the cosmic-ray ionization rate to be $\zeta_0=1.3\times10^{-17} \rm{s^{-1}}$ (Caselli et al. 1998). The mass-to-flux ratio (M/$\Upphi$) is given in units of the critical value for collapse (Mouschovias \& Spitzer 1976). The visual extinction ($A_v$) and temperature (T) are constant in each model. In model \texttt{M/$\Phi$0.5\_$\zeta/\zeta_01$\_$A_v$10\_T10\_dC} we leave the chemistry to evolve at our initial number density of $n_{\rm{H_2}} = 300 ~\rm{cm^{-3}}$ for 2 Myr before switching on gravity.}
\end{table}

\subsection{Theoretical results}\label{dynares}

\subsubsection{Time Evolution}\label{timeevol}

\begin{figure*}
\includegraphics[width=2.1\columnwidth, clip]{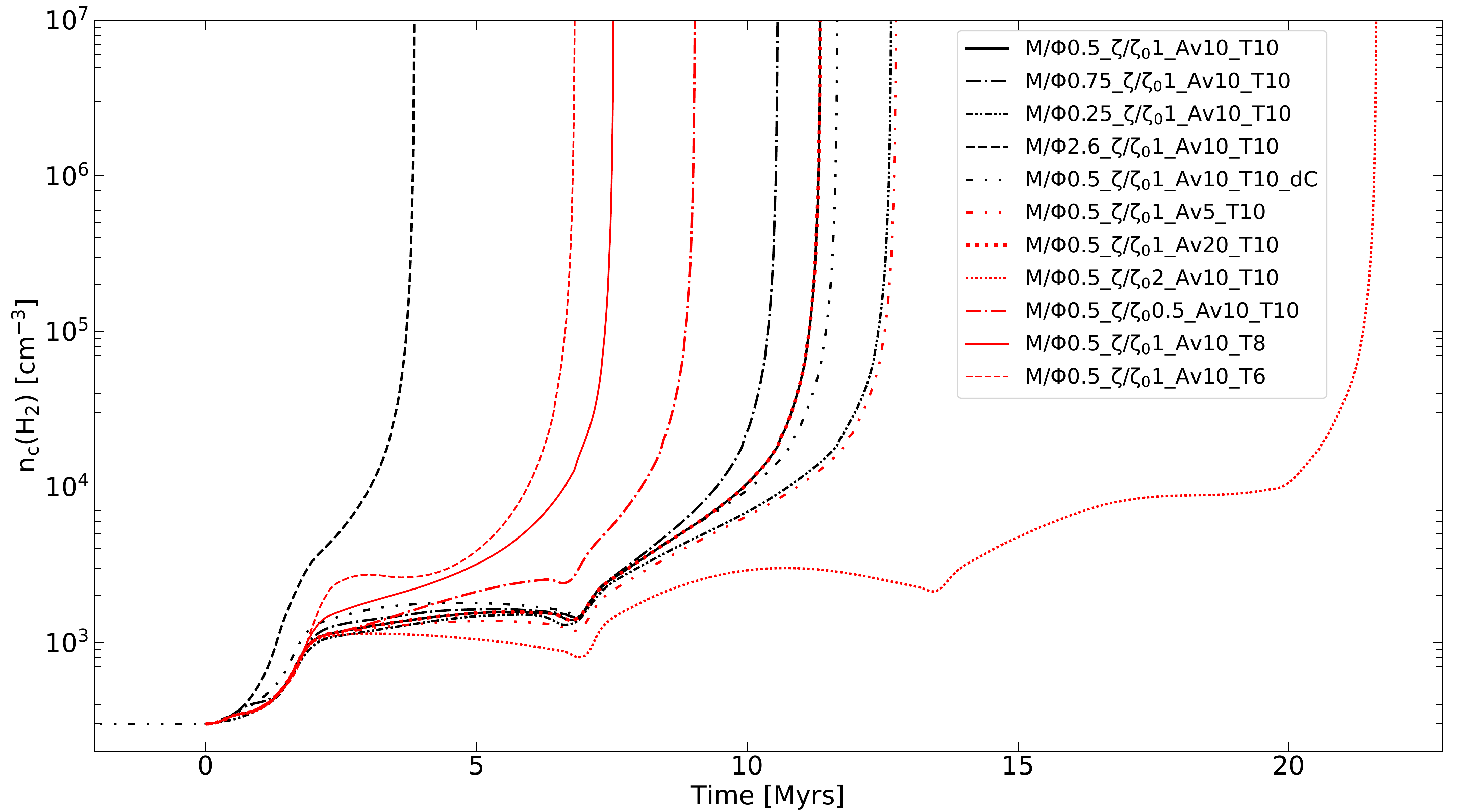}
\caption{Evolution of the central density as a function of time for all of our simulated cores. The physical parameters of each model are listed in Table~\ref{simstable}.
\label{CdensEvol}}
\end{figure*}

In Fig.~\ref{CdensEvol} we show the evolution of the central density as a function of time for each of our simulated clouds. Fig.~\ref{CdensEvol} reveals that the time evolution of the clouds is degenerate in terms of their initial mass-to-flux ratio. Specifically, it can be seen that a cloud with an initial mass-to-flux ratio of 0.5 (solid black line) can collapse \textit{faster} than a cloud with an initial mass-to-flux ratio of 0.75 (black dashed-dotted line) if its cosmic-ray ionization rate and/or temperature is lower (red dashed-dotted, solid and dashed lines, respectively).

Comparing the results from our fiducial model (solid black line; \texttt{M/$\Phi$0.5\_$\zeta/\zeta_01$\_$A_v$10\_T10}) with the rest of the models it is evident that a decrease in temperature by a few degrees Kelvin can reduce the ambipolar-diffusion timescale by almost a factor of two (solid and dashed red lines; \texttt{M/$\Phi$0.5\_$\zeta/\zeta_01$\_$A_v$10\_T8} and \texttt{M/$\Phi$0.5\_$\zeta/\zeta_01$\_$A_v$10\_T6}). Rather surprisingly, reducing the temperature has a greater effect than reducing the cosmic-ray ionization rate by a factor of two (see red dash-dotted line; \texttt{M/$\Phi$0.5\_$\zeta/\zeta_00.5$\_$A_v$10\_T10}). The reason behind this significant effect of reducing the temperature is twofold. First, at central densities of $n_{\rm{H_2}} \approx 10^3 - 3\times10^3 ~\rm{cm^{-3}}$, the number density of $\rm{H_3^+}$ which has been shown to be the dominant species carrying the perpendicular conductivity (Tassis et al. 2012a; Tritsis et al. 2022) can be an order of magnitude less in model \texttt{M/$\Phi$0.5\_$\zeta/\zeta_01$\_$A_v$10\_T6} compared to model \texttt{M/$\Phi$0.5\_$\zeta/\zeta_01$\_$A_v$10\_T10}. More importantly however, in the two models where the temperature is lower, after the initial thermal relaxation phase along magnetic field lines, the density is higher. The extra compression allowed by the reduction of temperature, also leads to more significant pinching of the magnetic field lines which in turn leads to a higher current density and drift velocities ($v_{\rm{dr}}\propto \eta_{\perp}J_{\phi}$; see Appendix~\ref{vdriftResiRelDeriv}).

From Fig.~\ref{CdensEvol} it is also evident that the cloud where the cosmic-ray ionization rate is two times higher than the standard value ($\zeta_0=1.3\times10^{-17} \rm{s^{-1}}$; Caselli et al. 1998) takes more than 20 Myr to collapse (red densely dotted line; \texttt{M/$\Phi$0.5\_$\zeta/\zeta_02$\_$A_v$10\_T10}). However, as we will show in \S~\ref{ChemEvolSec}, this model leads to very small abundances for a number of molecules (especially nitrogen-bearing ones) which are inconsistent with radio observations of clouds and cores. We therefore argue that this part of the parameter space is rejected by observations. On the other hand, the cloud where the cosmic-ray ionization rate is half the standard value (red dashed-dotted line; \texttt{M/$\Phi$0.5\_$\zeta/\zeta_00.5$\_$A_v$10\_T10}) collapses $\sim2.4$ Myr faster than our fiducial model and $\sim1.5$ Myr faster than the model where the mass-to-flux ratio is 0.75. 

Unlike the cosmic-ray ionization rate, the effect of visual extinction is nontrivial. That is, reducing the visual extinction by a factor of two (red loosely dashed-dot-dotted line; \texttt{M/$\Phi$0.5\_$\zeta/\zeta_01$\_$A_v$5\_T10}) delays the collapse by $\sim1$ Myr in comparison to our fiducial model. However, increasing the $A_v$ by a factor of two (red loosely dotted line; \texttt{M/$\Phi$0.5\_$\zeta/\zeta_01$\_$A_v$20\_T10}) makes no difference in the evolution of the cloud in comparison to our fiducial model.

Finally, leaving the chemistry to evolve for 2 Myr prior to switching on the gravity solver and allowing the cloud to evolve (black loosely dashed-dot-dotted line; \texttt{M/$\Phi$0.5\_$\zeta/\zeta_01$\_$A_v$10\_T10\_dC}) does not significantly affect the dynamical evolution of the cloud. In Fig.~\ref{CdensEvol} the line showing the evolution of model \texttt{M/$\Phi$0.5\_$\zeta/\zeta_01$\_$A_v$10\_T10\_dC} is shifted such that $t=0\ \rm{Myr}$ corresponds to the time we switch on the gravity solver. Additionally, as we will show in \S~\ref{ChemEvolSec}, this model leads to very similar chemical abundances as in our fiducial model.

\subsubsection{Spatial Comparison}\label{spatcomp}

\begin{figure*}
\includegraphics[width=2.1\columnwidth, clip]{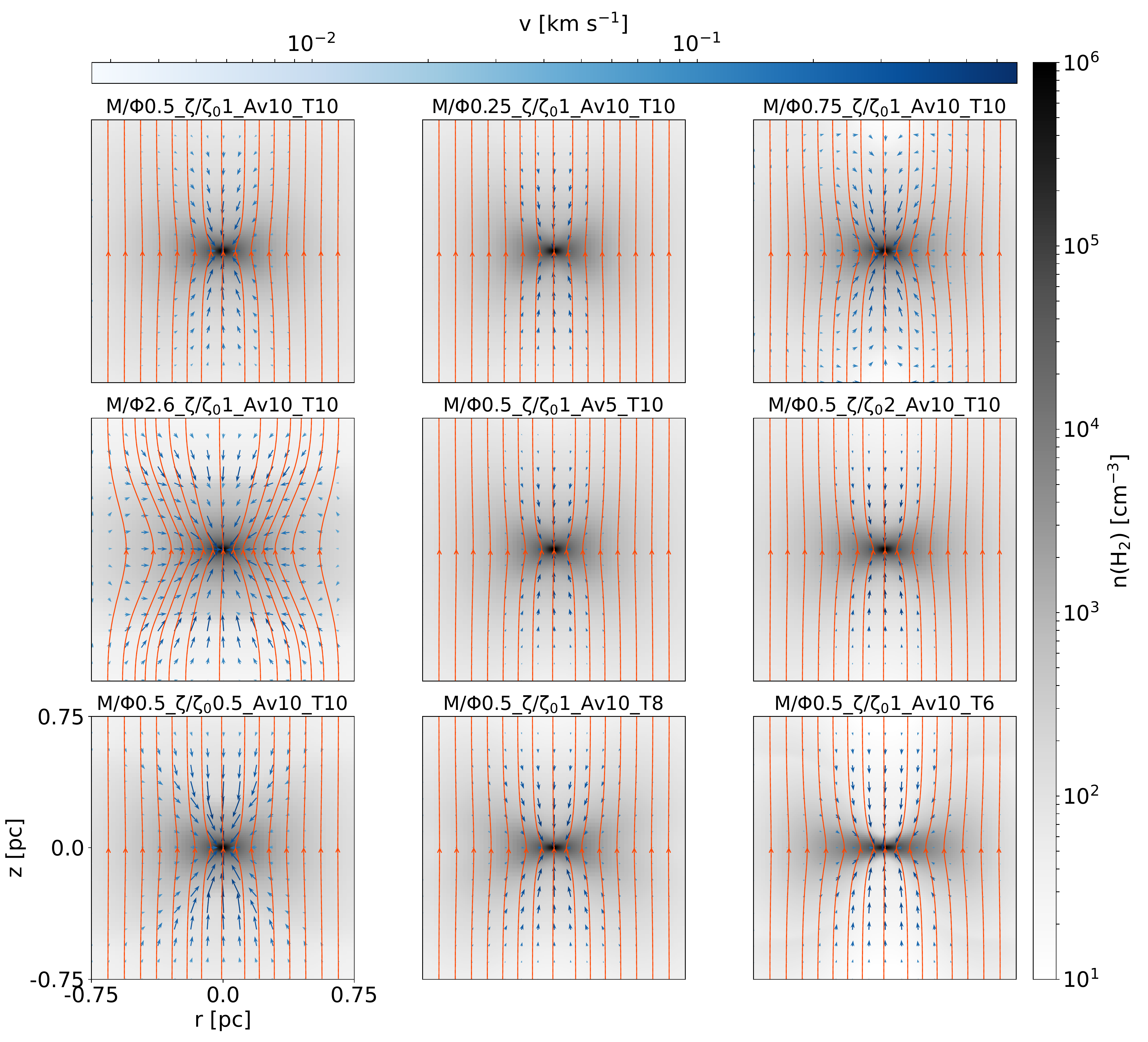}
\caption{Comparison of the spatial structure between our simulated cores when the central density is $n_{\rm{H_2}} = 10^6 ~\rm{cm^{-3}}$. Each panel is titled based on the corresponding model name. Results from models \texttt{M/$\Phi$0.5\_$\zeta/\zeta_01$\_$A_v$20\_T10} and \texttt{M/$\Phi$0.5\_$\zeta/\zeta_01$\_$A_v$10\_T10\_dC} are not shown since these models do not exhibit any significant differences in comparison to our fiducial model \texttt{M/$\Phi$0.5\_$\zeta/\zeta_01$\_$A_v$10\_T10}. In each panel, the orange streamlines show the magnetic field lines and the colour-coded blue arrows show the velocity vectors.
\label{SpatialComp}}
\end{figure*}

In Fig.~\ref{SpatialComp} we compare the spatial structure of our simulated clouds when the central density of all cores is $n_{\rm{H_2}}=10^6~\rm{cm^{-3}}$. At this evolutionary stage, the mass of the resulting core in our fiducial core, assuming a radius of 0.025 pc, is $\sim$1.4 solar masses which corresponds well to observations of nearby prestellar cores such as L1544 (e.g. Tafalla et al. 1998). We therefore stress here that not all the mass initially in our simulation region is accreted onto the core.

With the exception of our supercritical model (\texttt{M/$\Phi$2.6\_$\zeta/\zeta_01$\_$A_v$10\_T10}; left panel in the middle row) and our models where the temperature is lower (\texttt{M/$\Phi$0.5\_$\zeta/\zeta_01$\_$A_v$10\_T8} and \texttt{M/$\Phi$0.5\_$\zeta/\zeta_01$\_$A_v$10\_T6}; two rightmost panels in the bottom row) the rest of the cores do not exhibit significant differences in terms of their spatial structure (i.e. density distribution). Interestingly enough however, the differences in the physical conditions of the clouds, such as the mass-to-flux ratio, are sometimes imprinted in the velocity structure of the cloud (see for instance right panel in the upper row; \texttt{M/$\Phi$0.75\_$\zeta/\zeta_01$\_$A_v$10\_T10} and left panel in the bottom row; \texttt{M/$\Phi$0.5\_$\zeta/\zeta_00.5$\_$A_v$10\_T10}). Therefore, even if we cannot break the degeneracy and probe the initial conditions of clouds and cores by studying their spatial structure, we may be able to do so by studying their kinematic properties through radio observations.

\subsubsection{Radial Profiles of the ion-neutral drift velocity}\label{driftvelSims}

\begin{figure}
\includegraphics[width=1.\columnwidth, clip]{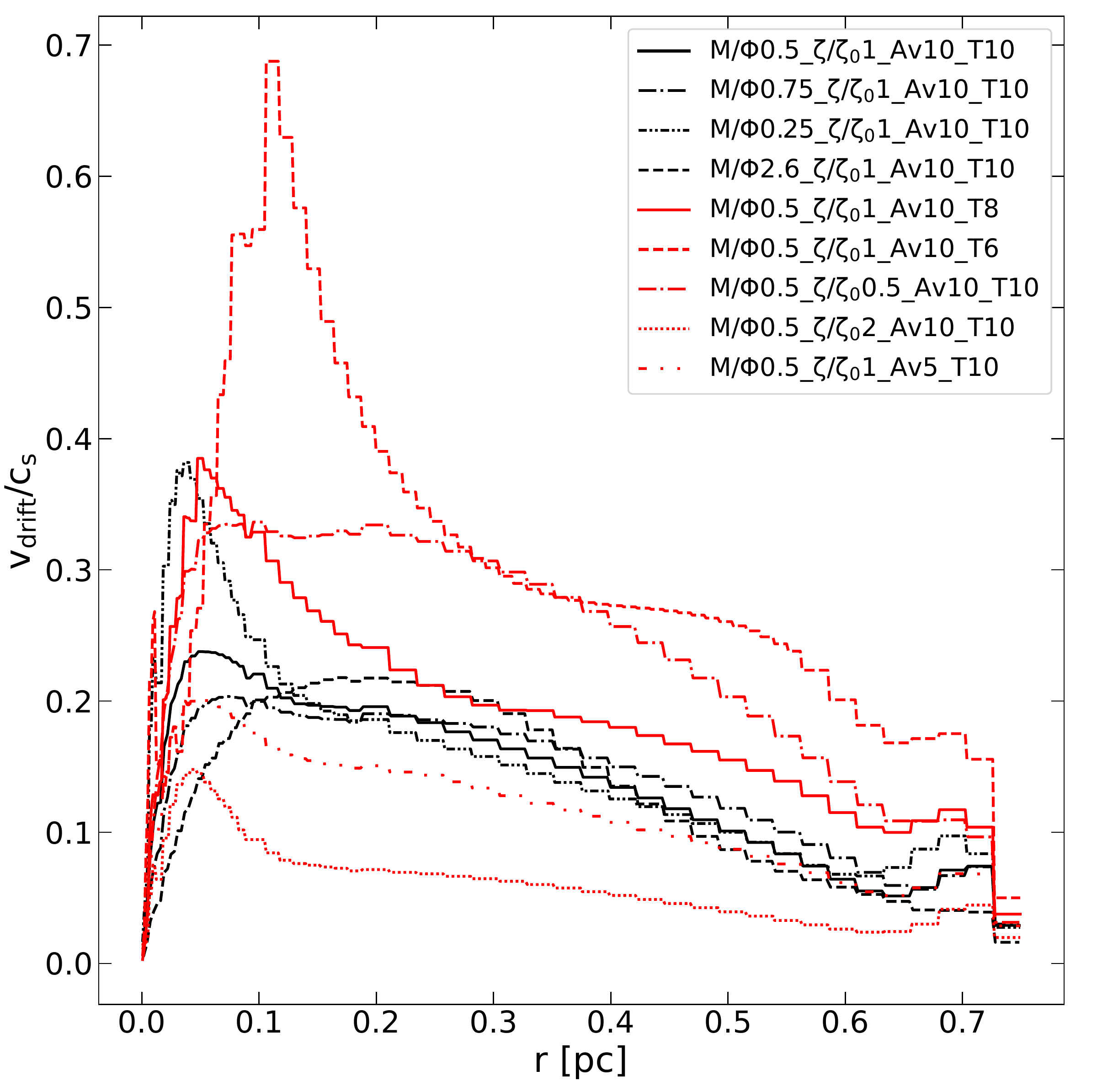}
\caption{Axially averaged (within $\pm$0.1 pc from the midplane) radial profiles of the ion-neutral drift velocity for each of our simulated cores when the central density is $n_{\rm{H_2}} = 10^6 ~\rm{cm^{-3}}$. Results from models \texttt{M/$\Phi$0.5\_$\zeta/\zeta_01$\_$A_v$20\_T10} and \texttt{M/$\Phi$0.5\_$\zeta/\zeta_01$\_$A_v$10\_T10\_dC} are again omitted since they are practically identical to our fiducial model \texttt{M/$\Phi$0.5\_$\zeta/\zeta_01$\_$A_v$10\_T10}.
\label{VdriftsRProfs}}
\end{figure}

In Fig.~\ref{VdriftsRProfs} we show the radial profiles of the ion-neutral drift velocity in units of the sound speed ($c_{\rm{s}} = 0.18, 0.16, 0.14 ~\rm{km~s^{-1}}$ for $T$ = 10, 8 and 6~\rm{K}, respectively) for each of our simulations when the central density of the clouds is $n_{\rm{H_2}} = 10^6 ~\rm{cm^{-3}}$. Each of the radial profiles presented was axially averaged within $\pm$0.1 pc from the midplane of the cloud. The drift velocity, defined as $\boldsymbol{v}_{\rm{dr}} = \boldsymbol{v}_{\rm{n}} - \boldsymbol{v}_{\rm{s}}$ (where $\boldsymbol{v}_{\rm{n}}$ is the velocity of the neutrals and $\boldsymbol{v}_{\rm{s}}$ is the velocity of species ``s"), is calculated as described in Eq.~\ref{driftVelEqs}. For the density range where ions remain well attached to the magnetic field, the drift velocity can also be physically understood as the difference in velocity between the neutrals and the velocity with which the magnetic field is advected (see for instance Eq. 20 from Tassis \& Mouschovias 2007).

Fig.~\ref{VdriftsRProfs} reveals very interesting features regarding the dependence of the drift velocity on the initial mass-to-flux ratio. Specifically, all our subcritical models (black solid, dashed-dotted and dashed-dot-dotted lines) exhibit the same profile shapes for the drift velocity. Surprisingly however, when the central density is $n_{\rm{H_2}} = 10^6 ~\rm{cm^{-3}}$, the model that exhibits the maximum drift velocity among all three models is the one with the smallest mass-to-flux ratio of 0.25 (dashed-dot-dotted lines). This behavior can be understood in terms of the chemical evolution of the cloud. Since the model with a mass-to-flux of 0.25 takes the longest to collapse among these three models, more ions are depleted from the gas phase onto dust grains where they can recombine. Consequently, when this cloud becomes supercritical, the neutrals are less coupled with the ions and the drift velocity is higher. Another interesting feature revealed from Fig.~\ref{VdriftsRProfs} is that the radial profile of the drift velocity in our supercritical cloud (black dashed line; \texttt{M/$\Phi$2.6\_$\zeta/\zeta_01$\_$A_v$10\_T10}) is much broader and peaks at larger radii in comparison to our subcritical models. However, the value of the drift velocity is no higher than that in our fiducial model.

As expected, changing the cosmic-ray ionization rate a factor of two above (red dotted line; \texttt{M/$\Phi$0.5\_$\zeta/\zeta_02$\_$A_v$10\_T10}) the standard value leads to lower drift velocities across all radii with the shape of the radial profile of the drift velocity changing only slightly when compared to our fiducial model (i.e. the peak in the inner $\sim$0.1 pc portion of the core is more prominent). The same argument can be made for the model where the visual extinction is set to 5 (red dashed-dot-dotted line; \texttt{M/$\Phi$0.5\_$\zeta/\zeta_01$\_$A_v$5\_T10}), a factor of two below our fiducial choice. However, decreasing the cosmic-ray ionization rate a factor of two below the standard value (red dashed-dotted line; \texttt{M/$\Phi$0.5\_$\zeta/\zeta_00.5$\_$A_v$10\_T10}) not only leads to an increase in the drift velocity across all radii, but also leads to a plateau of high drift velocities from $r\approx0.05$ to $r\approx0.3$ pc.

Surprisingly enough however, the parameter that has the greatest effect on the amplitude of the drift velocity is not the cosmic-ray ionization rate but is rather the temperature. Specifically, changing the temperature by just four degrees Kelvin, compared to our fiducial model, leads to a peak in the drift velocity that is almost comparable to the sound speed ($v_{\rm{dr}}^{\rm{Theoretical}}\approx0.7~c_{\rm{s}}$ see red dotted line; \texttt{M/$\Phi$0.5\_$\zeta/\zeta_00.5$\_$A_v$10\_T6}). Therefore, if we wish to maximize our chances of observing gravitationally-driven ambipolar diffusion in prestellar cores, \textit{we should be targeting the coldest cores}.

We note here that at a central number density of $n_{\rm{H_2}} = 10^6 ~\rm{cm^{-3}}$, the maximum difference in the drift velocity between our fiducial model and an identical model with an MRN grain distribution (Mathis et al. 1977; see simulations by Tritsis et al. 2022) is located at $r\sim0.05~\rm{pc}$ (i.e. where the drift velocity peaks for most models in Fig.~\ref{VdriftsRProfs}) and is equal to $0.012 ~\rm{km~s^{-1}}$. At larger radii the difference in the drift velocity between these two models is typically of the order of $0.003 ~\rm{km~s^{-1}}$. These small differences can be easily realized by studying the bottom left panel of Figure 4 from Tritsis et al. (2022) which shows that the differences in the ambipolar diffusion resistivity at this central number density are also minimal. In contrast, such differences are significantly smaller than the differences in drift velocity we typically observe when we alter other physical parameters.


\section{Radiative-transfer Simulations}\label{RT}

We use the \textsc{PyRaTE} (Tritsis et al. 2018) line radiative-transfer code to post-process our chemodynamical simulations presented in \S~\ref{dyna}. \textsc{PyRaTE} is a non-local thermodynamic equilibrium (non-LTE) radiative-transfer, multilevel code. The population densities of the species are calculated using the escape probability approach, taking into account variations in density, velocity and molecular number density (and temperature, if any). Specifically, for a grid point under consideration, we begin by solving the statistical equilibrium equations under LTE conditions (i.e. $\beta_{J_{\rm{u}}\rightarrow J_{\rm{l}}} = 1$, where $J_{\rm{u}}$ and $J_{\rm{l}}$ denote the ``upper" and ``lower" rotational energy levels). With these newly computed population densities, we compute the infinitesimal optical depth and compute the total optical depth by adding the infinitesimal optical depth of all grid points for which their velocity difference with the grid point under consideration is smaller than the thermal linewidth. The optical depth is calculated for six rays across the principal axes of the grid in Cartesian geometry (six-ray approximation) or four rays in cylindrical geometry. We then select the minimum optical depth from all rays under the assumption that it will be easier for the photons to escape along that direction, and compute new escape probabilities for each transition between the rotational energy levels. Finally, the process continues iteratively until the population densities of all rotational levels simultaneously satisfy a converge tolerance. For the purposes of the numerical experiments presented below we focus on model \texttt{M/$\Phi$0.5\_$\zeta/\zeta_01$\_$A_v$10\_T10} when the central density is $n_{\rm{H_2}} = 10^6 ~\rm{cm^{-3}}$.

All of our radiative-transfer calculations are performed using a spectral resolution of 0.01 $\rm{km~s^{-1}}$, considering five energy levels and assuming an edge-on view of the cloud. For our calculations, we use collisional and Einstein coefficients from the \textsc{LAMBDA} database (Sch{\"o}ier et al. 2005). Additionally, all of our calculations are performed at the highest refinement level of the non-ideal MHD simulations. When the central density is $n_{\rm{H_2}} = 10^6 ~\rm{cm^{-3}}$, this corresponds to 2.9$\times 10^{-3}~\rm{pc}$. However, to save computational time, we zoom-in to the inner 50\% of the cloud.

\begin{figure}
\includegraphics[width=1.\columnwidth, clip]{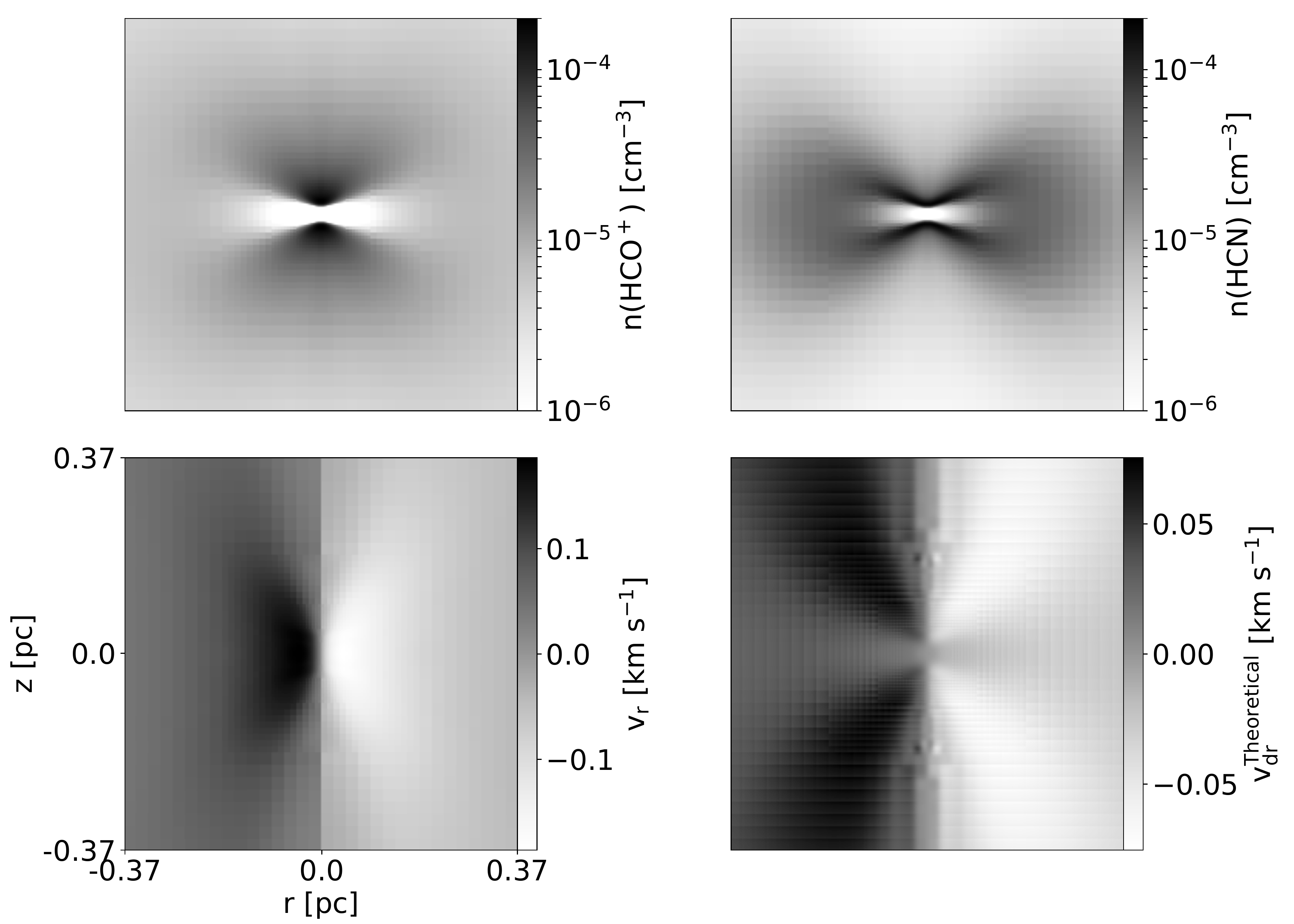}
\caption{Input parameters to \textsc{PyRaTE}. Upper left: number density of $\rm{HCO^+}$. Upper right: number density of $\rm{HCN}$. Lower left: radial velocity component. Lower right: drift velocity in the radial direction.
\label{pyrateinpts}}
\end{figure}

The inputs to \textsc{PyRaTE} are shown in Fig~\ref{pyrateinpts}. The upper left and right panels show the number density distribution of $\rm{HCO^+}$ and $\rm{HCN}$, respectively when the central density is $n_{\rm{H_2}} = 10^6 ~\rm{cm^{-3}}$. In the lower left and right panels we show the radial component of the velocity and the radial component of the drift velocity, respectively. The $\rm{H_2}$ number density is already shown in the upper left panel of ~\ref{SpatialComp}.

\subsection{Perfectly co-evolving species in LTE}\label{LTE}

In this numerical experiment we explore the simplest possible scenario; that of two perfectly chemically co-evolving molecular species emitting under LTE conditions. The aim of this numerical experiment is to verify whether in the simplest possible case, the drift velocity can ``survive" radiative-transfer effects and be detected. To this end, we first perform our radiative-transfer calculations considering that $\rm{HCO^+}$ moves with the velocity of the ions. We then consider that $\rm{HCO^+}$ moves with the velocity of the neutrals. This choice might at first seem strange. However, it is an indisputable way to isolate radiative-transfer from chemical effects, since no two molecules in our chemodynamical simulations are perfectly co-evolving. The reason behind our choice of $\rm{HCO^+}$ for our numerical experiments is that it has been used in previous observational studies of the ion-neutral drift velocity and it is a typical high-density tracing charged species.

\begin{figure}
\includegraphics[width=1.01\columnwidth, clip]{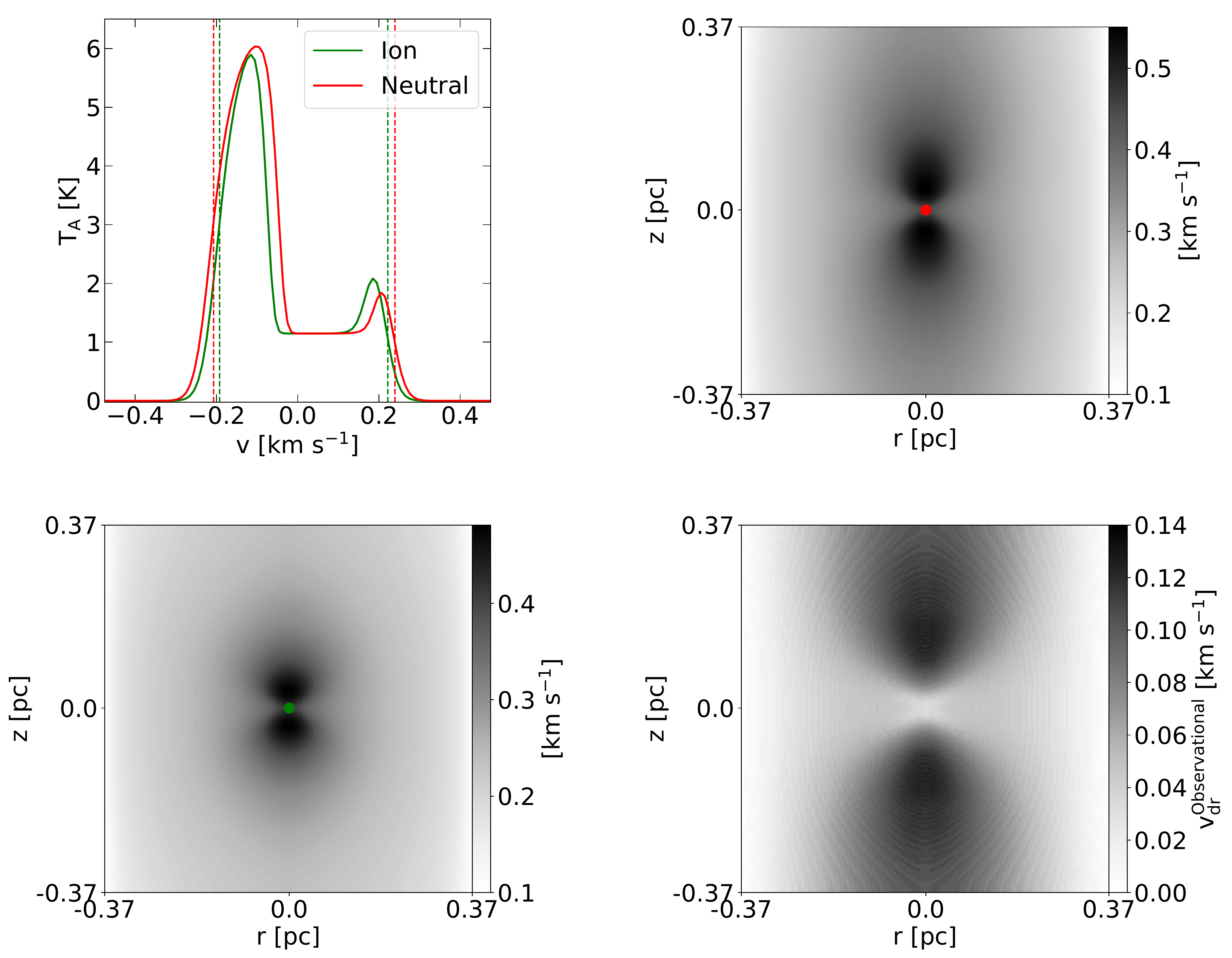}
\caption{Upper left: Simulated spectra of $\rm{HCO^+}$ under LTE conditions at the center of the core (red and green dots in the upper right and lower left panels) when assumed to be moving with the velocity of the neutrals (red line) and the velocity of the ions (green line). Upper right: Mock second moment map when $\rm{HCO^+}$ is taken to move with the velocity of the neutrals. Lower left: Mock second moment map when $\rm{HCO^+}$ is taken to move with the velocity of the ions. Lower right: observationally-derived map of the drift velocity.
\label{ltefig}}
\end{figure}

In the upper left panel of Fig.~\ref{ltefig}, we show two spectra of the $\rm{HCO^+}$ (\textit{J} = 1$\rightarrow$0) transition ($\nu_0 = 89.189~\rm{GHz}$) under LTE conditions where, in one instance, we take $\rm{HCO^+}$ to move with the velocity of the ions (green line) whereas in the other instance, we assumed that $\rm{HCO^+}$ is moving with the velocity of the neutrals (red line). We then measure the full width at half maximum (FWHM) of such spectra in each spatial position of the core when $\rm{HCO^+}$ is taken to move with the velocity of the neutrals and the ions (upper right and lower left panels, respectively). With the dashed lines in the upper left panel we denote the region that we consider to correspond to the FWHM (that is the FWHM of both the red and blue-shifted components). In order to avoid sharp changes in the FWHM that we measure due to the finite spectral resolution, we first perform linear interpolation of the spectra. Finally, in the bottom right panel we show the observationally-derived map of the drift velocity.

At first glance, a comparison of the lower left panel of Fig.~\ref{ltefig} with the lower left panel of Fig.~\ref{pyrateinpts}, yields significant differences. In reality however, the observationally-derived map of the drift velocity corresponds well to the theoretical drift velocity (lower right panel in Fig.~\ref{pyrateinpts}), as this is directly calculated from our non-ideal MHD chemodynamical simulations. Specifically, the theoretical drift velocity in the core contributes two times in the spectrum (from $r\geq0$ and $r\leq0$) and, as a result, the maximum value of the observationally-derived drift velocity at $r=0$ and $z\neq0$ (i.e. the axis of symmetry of the core) is $\sim$ twice the value of the theoretical one (e.g. $v_{\rm{dr, max}}^{\rm{Observational}}\approx0.14~\rm{km~s^{-1}}$ as opposed to $\lvert v_{\rm{dr, max}}^{\rm{Theoretical}}\rvert\approx0.07~\rm{km~s^{-1}}$). Additionally, the minimum value of the observationally-derived drift velocity is found in the midplane of the core, exactly as in the simulations. Thus, the drift velocity ``survives" radiative-transfer effects and can in principle be observed. At large radii, the observationally-derived drift velocity quickly approaches zero since, under axial symmetry, the line-of-sight component of the radial velocity becomes successively smaller. Therefore, given the finite spectral resolution, no difference between the two spectra can be measured. For more details on the comparison between the theoretical and observationally-derived maps of the drift velocity we refer to Appendix~\ref{VdriftProjected}.

\subsection{Perfectly co-evolving species in non-LTE}\label{nonLTE}

\begin{figure*}
\includegraphics[width=2.1\columnwidth, clip]{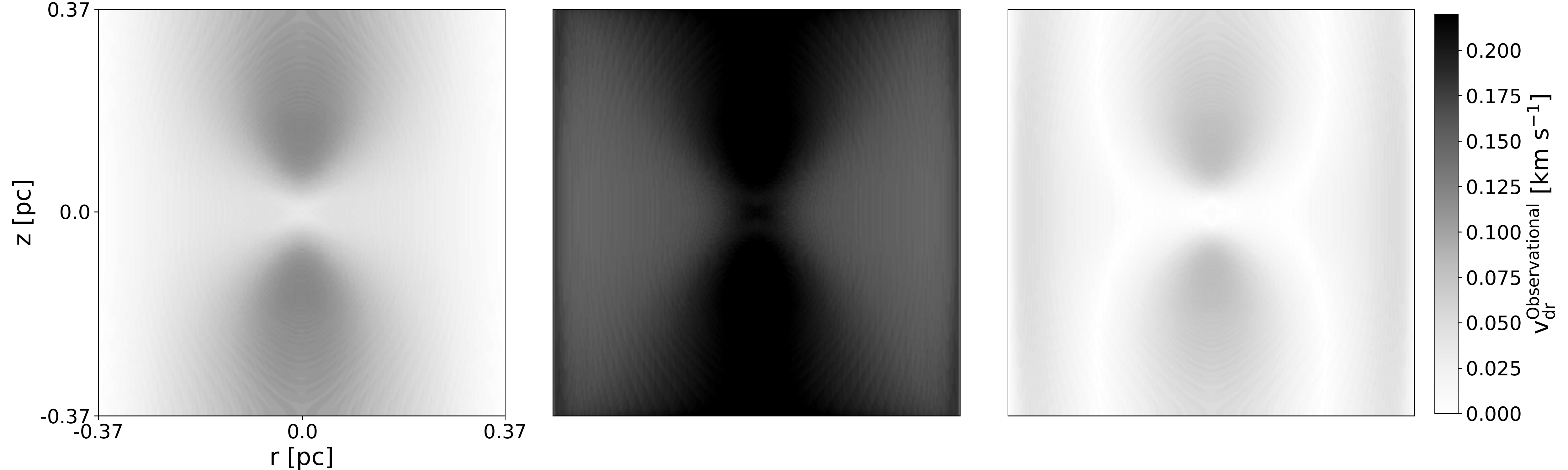}
\caption{Observationally-derived maps of the drift velocity in the case of two perfectly chemically co-evolving molecules emitting under non-LTE conditions (left; see \S~\ref{nonLTE} for details), in the case where the neutral species is taken to be very optically thick (middle; see \S~\ref{nonLTEthick}) and in the case where different collisional coefficients are assumed (right; see \S~\ref{h2cocoeffs}).
\label{nonLTEtests}}
\end{figure*}

In this section we further push the envelope by repeating the calculations presented above but in this instance we assume that $\rm{HCO^+}$ is emitting under non-LTE conditions. Non-LTE conditions are assumed regardless of whether $\rm{HCO^+}$ is taken to move with the velocity of the neutrals or that of the ions. The observationally-derived map of the drift velocity in this numerical experiment is shown in the left panel of Fig.~\ref{nonLTEtests}. While some minor differences are observed when comparing this result with the LTE case (lower right panel of Fig.~\ref{ltefig}), we argue that again the drift velocity ``survives" radiative-transfer effects. We also note here that we have repeated this numerical experiment when the central number density in our fiducial model is $n_{\rm{H_2}} = 10^5 ~\rm{cm^{-3}}$ and our results remain qualitatively unaltered.

\subsection{Perfectly co-evolving species in non-LTE; optically thick}\label{nonLTEthick}

To further explore radiative transfer effects we again repeat the calculations described in \S~\ref{nonLTE}, this time multiplying the number density of $\rm{HCO^+}$ by a factor of a thousand (when assumed to move with the velocity of the neutrals) such that it becomes very optically thick. To probe the velocity of ions we use our calculations from the previous section. In this manner, we also take into account a scenario where one of the molecules used to probe ambipolar diffusion is optically thin and the other is optically thick. The observationally-derived map of the drift velocity under such conditions is shown in the middle panel of Fig.~\ref{nonLTEtests}. Here, the value of the observationally-derived map of the drift velocity is consistently overestimated throughout the core. However, in terms of its qualitative structure, the observationally-derived map is also in fair agreement with the theoretical one.

\subsection{Linear molecule vs asymmetric top}\label{h2cocoeffs}

The collisional coefficients of each species with $\rm{H_2}$ depend on its internal structure. For instance, $\rm{HCO^+}$ is a linear molecule and will interact with $\rm{H_2}$ in a different manner than a non-linear species. As a result, a scenario can be realized where two species might be chemically co-evolving (and/or optically thin) but due to their collisional coefficients, radiative-transfer effects prevent us from probing ambipolar diffusion. Therefore, in this section, we take $\rm{HCO^+}$ to move with the velocity of the neutrals (and emit under non-LTE conditions) but use the collisional and Einstein coefficients from $\rm{H_2CO}$ which is an asymmetric molecule. For this numerical experiment we also use the frequencies (and hence energies of different rotational levels) of $\rm{H_2CO}$. However, we have verified that if we use the frequencies of $\rm{HCO^+}$ the resulting second moment map does not differ significantly, only having a small and almost constant offset (i.e. $0.012\pm 0.003~\rm{km ~s^{-1}}$) when compared to the second moment maps computed using the frequencies of $\rm{H_2CO}$.

To probe the velocity of the ions we again use our calculations from \S~\ref{nonLTE}. The observationally-derived map of the drift velocity under this scenario is shown in the right panel of Fig.~\ref{nonLTEtests}. Surprisingly, this is also a scenario where we observe notable differences. Specifically, the drop in the amplitude of the drift velocity towards the midplane of the core is more prominent towards the midplane of the core and at intermediate radii ($r\sim 0.18$ pc) while there is a further small increase in the amplitude of the drift velocity towards the edges of the cloud. However, even in this scenario, the observationally-derived drift velocity corresponds adequately well, in terms of the qualitative structures, to the theoretical radial drift velocity.

Therefore, based on these numerical experiments, \textit{we conclude that radiative-transfer effects do not significantly affect our ability to observe the ion-neutral drift velocity in molecular clouds}.

\begin{figure*}
\includegraphics[width=2.125\columnwidth, clip]{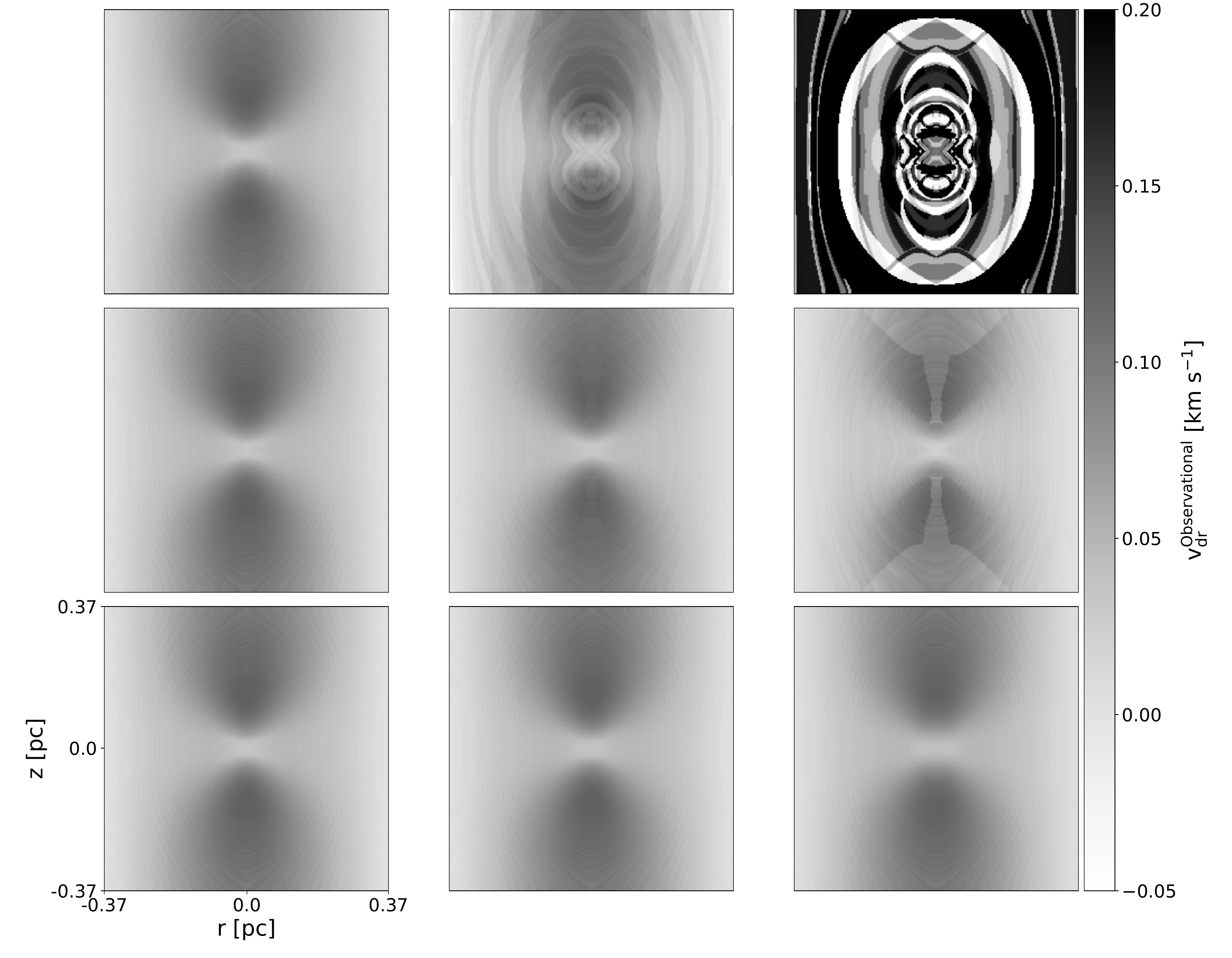}
\caption{Upper row: Observationally-derived maps of the drift velocity when decreasing the signal-to-noise ratio from 40 (left panel) to 20 (middle panel) and 10 (right panel). Middle row: Observationally-derived maps of the drift velocity gradually decreasing the spectral resolution from 0.025 (left panel) to 0.0625 (middle panel) and $\sim$0.156 $\rm{km~s^{-1}}$ (right panel). Bottom row: Same as the middle row but this time reducing the spatial resolution by the same factor as the spectral resolution. 
\label{ObsChlng}}
\end{figure*}

\subsection{Observational challenges}\label{observs}

Clearly, when it comes to probing the ion-neutral drift velocity in molecular clouds observational limitations are challenging to overcome. Especially, the spectral resolution we can achieve is of great importance when trying to observe velocity differences which are typically of the order of $\sim$20\%--30\% of the sound speed (see Fig.~\ref{VdriftsRProfs}). Using the Nobeyama 45 m Radio Telescope, Friesen et al. (2010) were able to perform $\rm{N_2H^+}$ observations towards the Ophiuchus B Core with an unprecedented spectral resolution of 0.025 $\rm{km~s^{-1}}$. Such a spectral resolution is only a factor of 2.5 greater than the spectral resolution of 0.01 $\rm{km~s^{-1}}$ assumed in our numerical experiments. Yet, it remains unclear whether this and other observational challenges could hinder a detection of the ion-neutral drift velocity. Therefore, in this section, we examine how the signal-to-noise ratio, spectral and spatial resolution can affect our ability to observe velocity differences in the spectra of charged and neutral species.

\subsubsection{signal-to-noise ratio}\label{snr}

We begin our exploration of observational challenges by examining the effect of the signal-to-noise ratio. We post-process our numerical calculations presented in \S~\ref{nonLTE}. We add Gaussian noise to the spectra in our simulated position-position-velocity data cubes so that we achieve a typical signal-to-noise ratio of 40, 20 and 10\footnote{Here, we draw caution to the fact that we do not impose these values for the signal-to-noise ratio for each spectrum in our simulated position-position-velocity cubes. Instead the quoted values are accurate for the spectra towards the center of the core while at the edges of the cloud, where the signal is much weaker, the signal-to-noise is much lower. Consequently, we essentially assume that both natural and artificial sources of interference are roughly constant ``during" our mock observations.}. We then calculate the second-moment maps and the corresponding observationally-derived maps of the drift velocity. Our results our shown in the upper row of Fig.~\ref{ObsChlng}.

From Fig.~\ref{ObsChlng}, it becomes apparent that \textit{a signal-to-noise ratio of at least 30--40 is required in order to robustly detect the ion-neutral drift velocity.}

\subsubsection{spectral and spatial resolution}\label{resol}

In this section, we explore the effects of spectral and spatial resolution in our ability to detect the ion-neutral drift velocity. For the spectral resolution we convolve each spectrum in our simulated position-position-velocity cubes presented in \S~\ref{nonLTE}, progressively reducing our resolution by a factor of 2.5 such that the worst spectral resolution is $\sim$0.16 $\rm{km~s^{-1}}$. Our results are shown in the middle row of Fig.~\ref{ObsChlng}. As it can be seen from Fig.~\ref{ObsChlng}, when the spectral resolution is of the order of $\sim$0.15 $\rm{km~s^{-1}}$ (typical of most present-day surveys), the observationally-derived map of the drift velocity does not correspond well to the theoretically-calculated drift velocity and exhibits additional ``artificial" features. 

Therefore, \textit{a spectral resolution of the order $\sim$0.05 $\rm{km~s^{-1}}$ (or less), is required for the ion-neutral drift velocity to be detected.} 

\smallskip

Similarly to the spectral resolution, we progressively reduce the spatial resolution by a factor of 2.5 by convolving our position-position-velocity cubes with a Gaussian kernel such that our worst spatial resolution is $\sim$0.045 pc (we remind the reader that at a central density of $n_{\rm{H_2}} = 10^6 ~\rm{cm^{-3}}$, our native spatial resolution is 2.9$\times 10^{-3}~\rm{pc}$). Our results are shown in the bottom row of Fig.~\ref{ObsChlng}. 

Based on the results presented in the bottom row of Fig.~\ref{ObsChlng}, \textit{we argue that spatial resolution does not significantly affect our ability of detecting the ion-neutral drift velocity}.

\subsection{Non co-evolving species: Drift velocity based on $\rm{HCN}$ and $\rm{HCO^+}$ mock observations}\label{reals}

Here, we explore a more realistic scenario where we compute the observationally-derived map of the ion-neutral drift velocity based on the $\rm{HCN}$ and $\rm{HCO^+}$ (\textit{J} = 1$\rightarrow$0) transitions under non-LTE conditions. The rest frequency of the $\rm{HCN}$ (\textit{J} = 1$\rightarrow$0) transition is 88.63 GHz. The number-density distribution of $\rm{HCN}$ in the core is shown in the upper right panel of Fig~\ref{pyrateinpts}. While this is aimed at being the most realistic situation we present in this study, we do not add any noise to the spectra, and/or convolve our simulated position-position-velocity cubes to reduce the spectral and spatial resolution in order to isolate chemical effects from any observational challenges already examined in \S~\ref{observs}.

\begin{figure}
\includegraphics[width=1.01\columnwidth, clip]{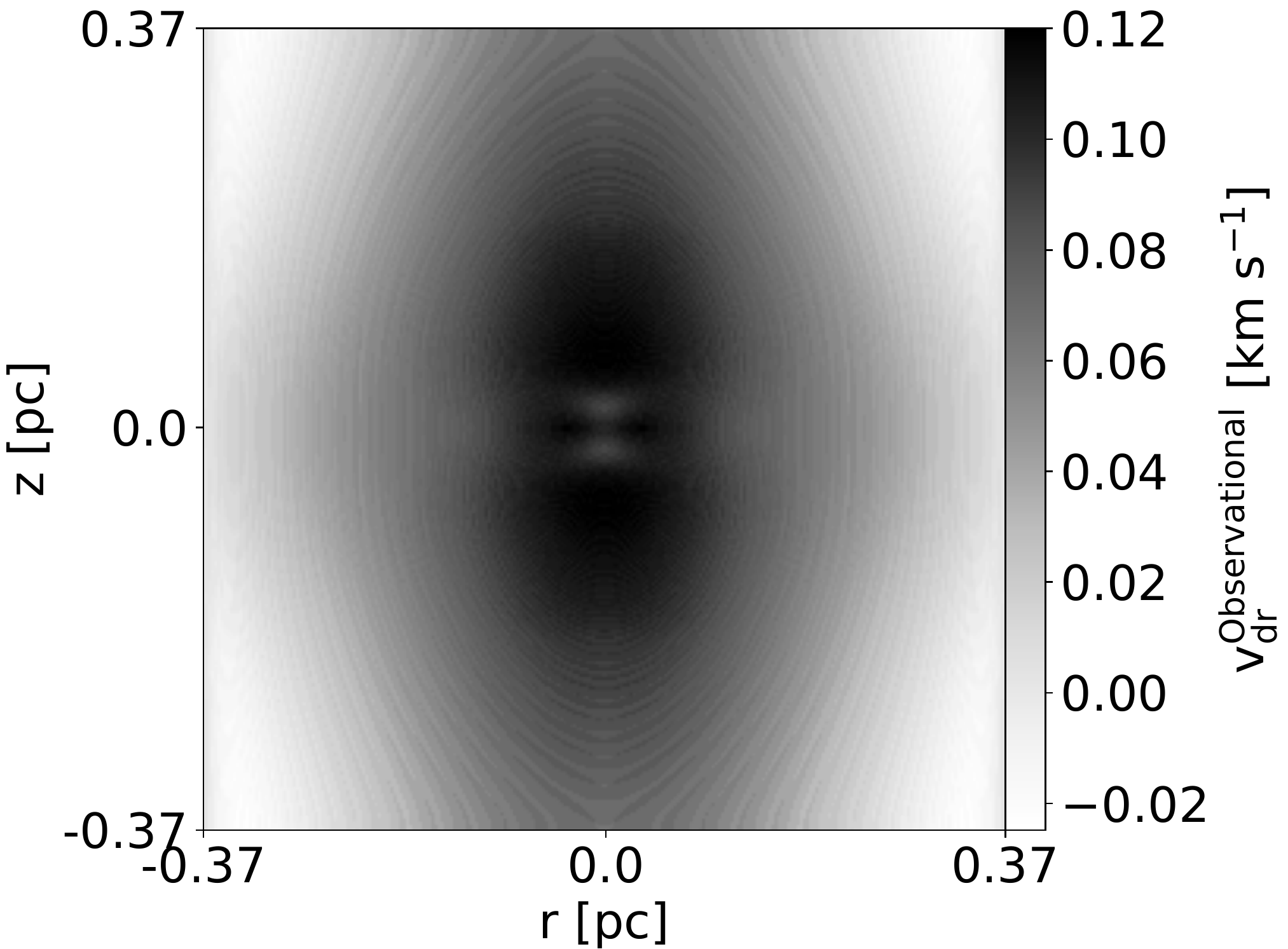}
\caption{Observationally-derived map of the drift velocity based on the $\rm{HCN}$ and $\rm{HCO^+}$ (\textit{J} = 1$\rightarrow$0) transitions under non-LTE conditions. 
\label{ChemEff}}
\end{figure}

The observationally-derived map of the ion-neutral drift velocity based on these species is presented in Fig.~\ref{ChemEff}. Clearly, when considering two non-chemically co-evolving species for computing the drift velocity the observationally-derived map no longer corresponds well to the theoretically calculated drift velocity. For instance, the flaring structure towards the midplane of the core where the theoretical drift velocity is close to zero (see bottom right panel in Fig.~\ref{pyrateinpts}) is lost. Instead, the observationally computed map of the drift velocity exhibits a more circular/elliptical pattern which cannot be justified based on the theoretical drift velocity. It therefore becomes clear that, even without any observational challenges taken into account, \textit{chemical effects have the most significant impact in our ability to robustly detect the ion-neutral drift velocity.}

\section{Chemical Evolution}\label{ChemEvolSec}

\begin{figure*}
\includegraphics[width=2.1\columnwidth, clip]{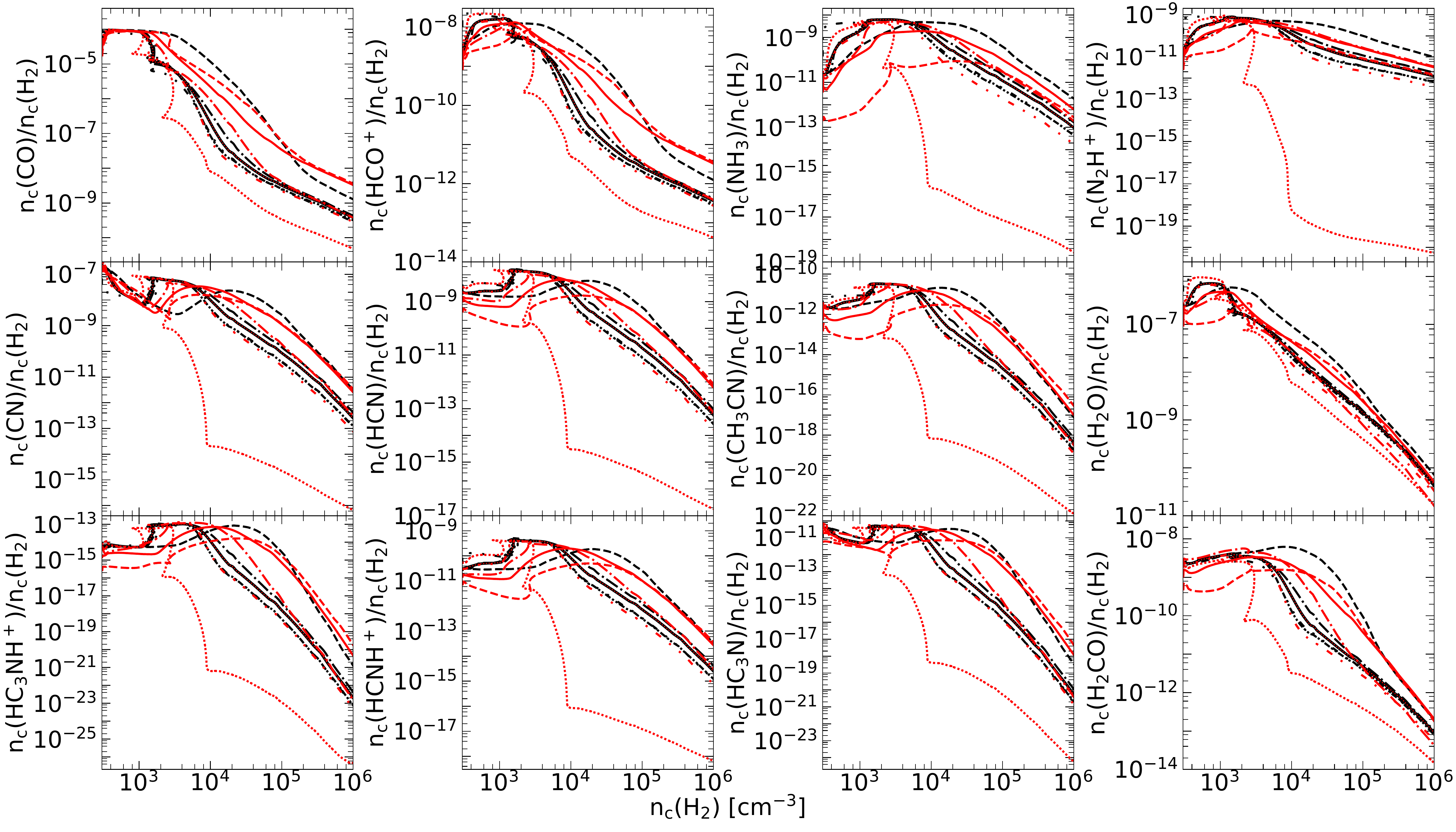}
\caption{Chemical evolution of various commonly observed neutral and charged species as a function of the central density of the cloud for each of our non-ideal MHD simulations. Linestyles are the same as in Figs.~\ref{CdensEvol} \&~\ref{VdriftsRProfs}. A pair of species that is co-evolving in all of our models is $\rm{HCN}$ (middle row, second panel) and $\rm{HCNH^+}$ (third row, second panel).
\label{ChemEvol}}
\end{figure*}

Given that chemical effects can significantly impede our view of the ion-neutral drift velocity we now seek to identify chemically co-evolving species in our simulations. In Fig.~\ref{ChemEvol} we show the abundance of various commonly observed molecules as a function of the central density of the core from all of our chemodynamical simulations presented in \S~\ref{dyna}.

One of the most striking features from Fig.~\ref{ChemEvol} is that in the simulation where the cosmic-ray ionization rate is two times higher than the standard value (red densely dotted line; \texttt{M/$\Phi$0.5\_$\zeta/\zeta_02$\_$A_v$10\_T10}) the abundances of various molecules (including ammonia and $\rm{N_2H^+}$) is extremely low. The reason behind these low abundances is that this model spends a significant portion of its evolution at low to intermediate densities, allowing for the depletion of these species onto dust grains. To our knowledge, no observational studies of prestellar cores have ever indicated such low abundances for neither ammonia nor $\rm{N_2H^+}$. Therefore, regardless of other observational constraints, this result indicates that subcritical cores with a high cosmic-ray ionization rate should be very uncommon in nature (if any exist at all). The same cannot be necessarily said for supercritical cores with a high cosmic-ray ionization rate as, due to the faster collapse, the abundances of these species may not be as low. However, more chemodynamical simulations are required to support the latter claim. Additionally, by examining Fig.~\ref{ChemEvol} a general argument can be made that chemistry acts as a ``clock" with most species exhibiting lower abundances (at the same central number density) for the modelled cores that collapse slower and higher abundances for the clouds that collapse faster. However, this effect is often highly degenerate. The chemical evolution of the cloud does not only depend on the physical parameters (e.g. temperature, cosmic-ray ionization rate, $A_v$) which are used as an input for calculating the reaction rates of the various chemical reactions but is also subject to the duration the cloud spends at different evolutionary stages. The time the cloud spends on the different stages is in turn determined by both the physical parameters, such as the mass-to-flux ratio, and the chemistry itself, since the resistivities which determine the ambipolar diffusion timescale, are determined by the chemistry. This interplay constitutes a complex feedback loop which can only be accurately modelled with numerical simulations such as the ones presented here.

From Fig.~\ref{ChemEvol} as well as the two upper panels of Fig~\ref{pyrateinpts}, it is also evident that \textit{$\rm{HCN}$ and $\rm{HCO^+}$, typically used in studies of the drift velocity are not chemically co-evolving}. The same statement also holds true for $\rm{NH_3}$ and $\rm{N_2H^+}$. This result is consistent with observations by Tafalla et al. (2002) who, based on a survey of five low-mass prestellar cores, concluded that the $\rm{N_2H^+}$ abundance remains roughly constant while the ammonia exhibits an increase towards the center of the core.

\begin{figure*}
\includegraphics[width=2.1\columnwidth, clip]{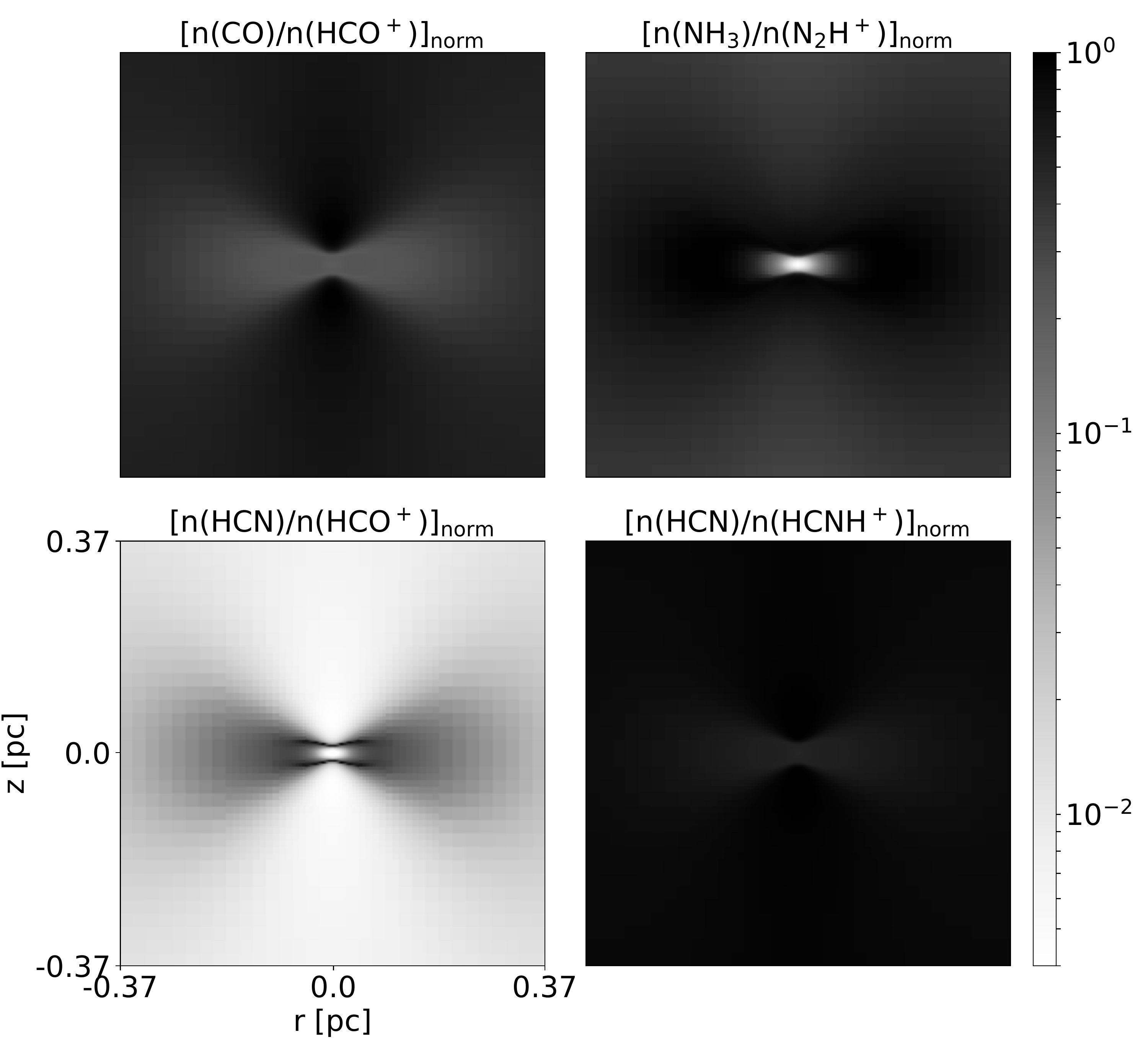}
\caption{Ratios between various pairs of charged and neutral species used (or proposed to be used) in observations of ambipolar diffusion. In each map the ratio is normalized to the corresponding maximum value. Clearly, the pair $\rm{HCN}$/$\rm{HCNH^+}$ (lower right) exhibits the least variation in the cloud.
\label{SpatialComp_MolRats}}
\end{figure*}

Based on Fig.~\ref{ChemEvol}, \textit{the species that are very similar in terms of their chemical evolution and could thus be used in future observational studies of the ion-neutral drift velocity are $\rm{HCN}$ (third row; second panel) and $\rm{HCNH^+}$ (fourth row; second panel)}. To further emphasize this point, in Fig.~\ref{SpatialComp_MolRats} we plot the ratios of various neutral to charged species. Each map is normalized to its maximum value since we are interested in the variation of each ratio rather than its absolute value. In the upper left panel we plot the ratio of $\rm{CO}$ to $\rm{HCO^+}$. In the upper right and lower left we plot the ratios of $\rm{NH_3}$/$\rm{N_2H^+}$ and $\rm{HCN}$/$\rm{HCO^+}$, respectively, which have already been used in such observations (see \S~\ref{intro}). Finally, in the lower right we plot the ratio of $\rm{HCN}$ to $\rm{HCNH^+}$, proposed herewith as a suitable species for probing the ion-neutral drift velocity. The ratio of $\rm{HCN}$ to $\rm{HCNH^+}$ exhibits a variation of just a factor of two in the cloud and is therefore the most promising candidate for future observations of ambipolar diffusion.

Based on their chemodynamical simulations, Tassis et al. (2012b) proposed that the pair of species $\rm{NO}$/$\rm{NO^+}$, $\rm{CO}$/$\rm{HCO^+}$ and/or $\rm{NO}$/$\rm{HCO^+}$ could be used to probe ambipolar diffusion in molecular clouds. Firstly, as we have shown in Fig.~\ref{SpatialComp_MolRats}, while the ratio of $\rm{CO}$ to $\rm{HCO^+}$ does not vary as much as others commonly used pairs of neutral/charged species, the pair $\rm{HCN}$/$\rm{HCNH^+}$ varies even less. Additionally, it is well known that $\rm{CO}$ is optically thick and thus observations based on $\rm{CO}$ would overestimate the ion-neutral drift velocity. Regarding the two other pairs proposed by Tassis et al. (2012b), while a tentative detection of $\rm{NO^+}$ has been reported in the literature (Cernicharo et al. 2014), $\rm{NO}$ has never been observed in the interstellar medium. On the other hand, $\rm{HCN}$ is a commonly observed species and has even be used in larger-scale surveys of molecular clouds (see for instance Storm et al. 2014; Lee et al. 2014 and Storm et al. 2016). The first detection of $\rm{HCNH^+}$ in the interstellar medium was reported by Ziurys \& Turner (1986). $\rm{HCNH^+}$ has since been observed in both low, and high-mass cores by Qu{\'e}nard et al. (2017) (and references therein) and Fontani et al. (2021), respectively. Additionally, both $\rm{HCN}$ and $\rm{HCNH^+}$ are linear molecules so that no significant complications should be expected in terms of their collisional coefficients with $\rm{H_2}$ (see discussion in \S~\ref{h2cocoeffs}). Unfortunately however, to our knowledge, no collisional and Einstein coefficients are available for $\rm{HCNH^+}$. Therefore, we cannot perform radiative-transfer calculations to properly test whether mock observations of these two species could accurately probe the ion-neutral drift velocity. 

\section{Summary \& Conclusions}\label{discuss}

We presented a suite of 2D axisymmetric, non-ideal MHD chemodynamical simulations of prestellar cores in order to study the ion-neutral drift velocity. We then performed a number of idealized and non-idealized, radiative-transfer calculations in order to explore whether the ion-neutral drift velocity could be observationally detected. Below we summarize our main conclusions:
\begin{itemize}
\item In the subcritical regime, the drift velocity is inversely correlated with the mass-to-flux ratio with more subcritical cores exhibiting a higher drift velocity at the same central density.
\item Temperature has a significant effect in the amplitude of the ion-neutral drift velocity with the coldest core (\textit{T} = 6 K) in our non-ideal MHD chemodynamical simulations exhibiting a maximum drift velocity in the radial direction of $\sim$70\% of the sound speed.
\item For the most part, radiative-transfer effects do not significantly impede our ability to detect the ion-neutral drift velocity.
\item In terms of observational challenges, a signal-to-noise ratio of $\sim$30--40 and a spectral resolution of the order of $\sim$0.05 $\rm{km~s^{-1}}$ would be required in order to observe the ion-neutral drift velocity. An exceptional spatial resolution does not appear to be a significant factor for such observational studies.
\item Chemical effects on the other hand are a deciding factor for observational studies of the gravitationally-initiated ambipolar diffusion. Specifically, using species that are not chemically co-evolving can lead to a skewed view of the ion-neutral drift velocity.
\item The pair of species $\rm{HCN}$ and $\rm{HCNH^+}$ is a very promising candidate for future observational studies of the ion-neutral drift velocity, as these two species were found to be chemically co-evolving in all our non-ideal MHD chemodynamical simulations.
\end{itemize}

\section*{Acknowledgements}

We thank K. Tassis for useful comments and discussions. We additionally thank the anonymous referee for suggestions that helped improve this manuscript. A. Tritsis acknowledges support by the Natural Sciences and Engineering Research Council of Canada (NSERC), [funding reference \#CITA 490888-16] and by the Ambizione grant no. PZ00P2\_202199 of the Swiss National Science Foundation (SNSF). C. Federrath acknowledges funding provided by the Australian Research Council (Future Fellowship FT180100495 and Discovery Projects DP230102280), and the Australia-Germany Joint Research Cooperation Scheme (UA-DAAD). This research was undertaken with the assistance of resources and services from the National Computational Infrastructure (NCI - grant ek9), supported by the Australian Government. The software used in this work was in part developed by the DOE NNSA-ASC OASCR Flash Center at the University of Chicago. We also acknowledge use of the following software: \textsc{Matplotlib} (Hunter 2007), \textsc{Numpy} (Harris et al. 2020) and the \textsc{yt} analysis toolkit (Turk et al. 2011).
 
\section*{DATA AVAILABILITY}

The non-ideal MHD chemodynamical simulations presented herewith are available at https://zenodo.org/record/7185442. The line-radiative transfer simulations are available from the corresponding author upon reasonable request.

\appendix
\section{Relation between the ion-neutral drift velocity and the resistivities}\label{vdriftResiRelDeriv}

The relation between the drift velocity and the resistivities can be derived by combining the momentum equation for charged species with the generalized Ohm's law:

\begin{equation}\label{momI}
0 = en_{\rm{s}}(\boldsymbol{E}+\frac{\boldsymbol{v}_{\rm{s}}}{c}\times\boldsymbol{B}) + \frac{\rho_{\rm{s}}}{\uptau_{\rm{sn}}}(\boldsymbol{v}_{\rm{n}} - \boldsymbol{v}_{\rm{s}})
\end{equation}

\begin{equation}\label{gohm}
\boldsymbol{E} + \frac{\boldsymbol{v}_{\rm{n}}}{c}\times\boldsymbol{B} = \eta_\perp \boldsymbol{j_\perp} + \eta_\parallel \boldsymbol{j_\parallel} + \eta_{\rm{H}}	\boldsymbol{j}\times \boldsymbol{b}
\end{equation}
where $n_{\rm{s}}$, $\rho_{\rm{s}}$ and $\boldsymbol{v}_{\rm{s}}$ denote, respectively, the number density, mass density and velocity of species ``s" and $\boldsymbol{v}_{\rm{n}}$ denotes the velocity of the neutrals. In Eq.~\ref{momI}, $\uptau_{\rm{sn}}$ is the mean collisional timescale between species s and the neutrals, and \textit{c} and \textit{e} denote the speed of light and electron charge, respectively. The quantities, $\boldsymbol{B}$ and $\boldsymbol{E}$ denote the magnetic and electric field, respectively, and $\boldsymbol{b}$ is the unit vector of the magnetic field. Finally, in Eq.~\ref{gohm}, $\boldsymbol{j_\perp}$, $\boldsymbol{j_\parallel}$ denote the current density perpendicular and parallel to the magnetic field, and $\eta_\perp$, $\eta_\parallel$ and $\eta_{\rm{H}}$ are the perpendicular, parallel and Hall resistivities. The mean collisional timescale is defined as:
\begin{equation}\label{mct}
\uptau_{\rm{sn}} = \frac{1}{\alpha_{\rm{sHe}}}\frac{m_{\rm{s}}+m_{\rm{H_2}}}{\rho_{\rm{H_2}}}\frac{1}{\mean{\sigma w}_{\rm{sH_2}}}
\end{equation}
where $\mean{\sigma w}_{\rm{sH_2}}$ is the mean collision rate which is computed from the Langevin approximation (e.g. Gioumousis \& Stevenson 1958). In Eq.~\ref{mct}, $m_{\rm{s}}$ and $m_{\rm{H_2}}$ denote the mass of species s and that of $\rm{H_2}$, $\rho_{\rm{H_2}}$ is the $\rm{H_2}$ mass density and $\alpha_{\rm{sHe}}$ is an additional factor to account for the slowing-down of species s due to the presence of $\rm{He}$.

Substituing the electric field in Eq.~\ref{momI} from Eq.~\ref{gohm}, defining $\boldsymbol{v}_{\rm{dr}} = \boldsymbol{v}_{\rm{n}} - \boldsymbol{v}_{\rm{s}}$, re-arranging and multiplying both sides with $c/en_{\rm{s}}$ we get:
\begin{equation}\label{resietarel}
\frac{m_{\rm{s}}c}{\uptau_{\rm{sn}}e}\boldsymbol{v}_{\rm{dr}} + \boldsymbol{v}_{\rm{dr}}\times\boldsymbol{B} = c(\eta_\perp \boldsymbol{j_\perp} + \eta_\parallel \boldsymbol{j_\parallel} + \eta_{\rm{H}}	\boldsymbol{j}\times \boldsymbol{b})
\end{equation}
which, in component form, can be written as:
\begin{eqnarray}\label{driftVelEqs}
\begin{bmatrix}
\frac{m_{\rm{s}}c}{\uptau_{\rm{sn}}e} & B_z & -B_y \\
-B_z & \frac{m_{\rm{s}}c}{\uptau_{\rm{sn}}e} & B_x \\
B_y & -B_x & \frac{m_{\rm{s}}c}{\uptau_{\rm{sn}}e}
\end{bmatrix}
\begin{bmatrix}
v_{{\rm{dr}}, x} \\
v_{{\rm{dr}}, y} \\
v_{{\rm{dr}}, z}
\end{bmatrix} = \\ \nonumber c
\begin{bmatrix}
\eta_\perp j_{\perp, x} + \eta_\parallel j_{\parallel, x} + \eta_{\rm{H}}(\boldsymbol{j}\times \boldsymbol{b})_x \\
\eta_\perp j_{\perp, y} + \eta_\parallel j_{\parallel, y} + \eta_{\rm{H}}(\boldsymbol{j}\times \boldsymbol{b})_y \\
\eta_\perp j_{\perp, z} + \eta_\parallel j_{\parallel, z} + \eta_{\rm{H}}(\boldsymbol{j}\times \boldsymbol{b})_z
\end{bmatrix}
\end{eqnarray}
which can be solved using Cramer's method to yield $v_{{\rm{dr}}, x}$, $v_{{\rm{dr}}, y}$ and $v_{{\rm{dr}}, z}$. We note here that in cylindrical geometry, the only non-vanishing component of the current density is $\boldsymbol{j_\perp}$ (i.e. $J_{\phi}$; Kunz \& Mouschovias 2009).


\section{Projected map of the theoretical ion-neutral drift velocity}\label{VdriftProjected}

While the observationally-derived map of the ion-neutral drift velocity (e.g. botton right panel of Fig.~\ref{ltefig}) can be qualitatively understood based on the theoretical one (bottom right panel of Fig.~\ref{pyrateinpts}), the two maps cannot be compared directly. The reason is that the theoretical map of the drift velocity essentially contains 3-dimensional information, while the observationally-derived map only contains 2D information.

\begin{figure}
\centering
\includegraphics[width=1.\columnwidth, clip]{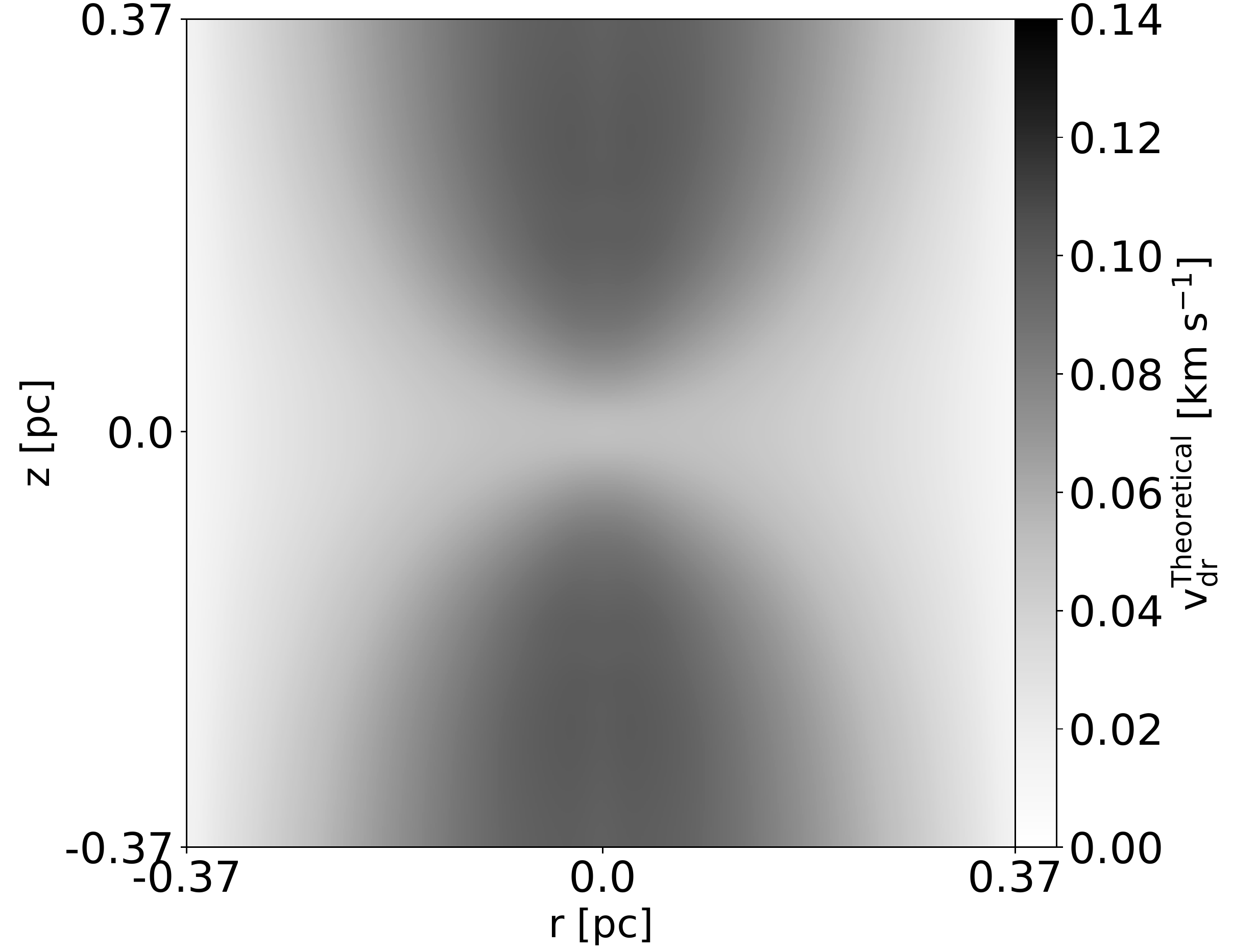}
\caption{Projected map of the LOS component of the ion-neutral drift velocity as this is computed in our chemodynamical simulations.
\label{TheoreticalVdrift_Projected}}
\end{figure}

Comparing the two maps requires a volume rendering of the absolute value of the theoretical map of the drift velocity and a ``projection" of its line-of-sight (LOS) component in each sightline. By the term ``projection" we signify that we consider twice the value of the mean LOS component of the ion-neutral drift velocity in each sightline. Considering the absolute value is required since in the opposite scenario the mean value would be zero everywhere (see lower right panel of Fig.~\ref{pyrateinpts}). The resulting, ``projected" map of the theoretical drift velocity is shown in Fig.~\ref{TheoreticalVdrift_Projected}.

We draw caution to the fact here that the ``projected" map of the theoretical drift velocity shown in Fig.~\ref{TheoreticalVdrift_Projected} is, to a certain extend, a simpler (and less accurate) version of some of our radiative transfer calculations (such as the case presented under \S~\ref{LTE}). The reason why this version is less accurate is that the resulting 2D map is not ``informed" by outliers in the value of ion-neutral drift velocity which however naturally appear in the spectra. If on the other hand we consider the maximum value instead of the mean for each sightline then the resulting map is only subject to one value.

Regardless of these limitations, the ``projected" map of the theoretical drift velocity is qualitatively very similar to the observationally-derived maps with differences mainly found at large radii where the line-of-sight component of the velocity becomes successively smaller and no observable difference can be measured due to the finite spectral resolution. Hence, observationally-derived maps of the ion-neutral drift velocity approach zero somewhat faster than the theoretical map shown in Fig.~\ref{TheoreticalVdrift_Projected}. Quantitatively, the mode of the distribution of the ratio of the observational (for the case presented under \S~\ref{LTE}) and the ``projected" theoretical maps is exactly unity. However, given the limitations in constructing the map presented in Fig.~\ref{TheoreticalVdrift_Projected} we do not perform a detailed quantitative comparison and instead focus on the qualitative picture.

\end{document}